\documentclass[12pt]{iopart}

\usepackage{graphicx}
\usepackage{epstopdf}
\usepackage[colorlinks=false]{hyperref}
\usepackage[all]{hypcap}
\usepackage{color}
\newcommand{\degree}{^{\circ}}

\begin{document}

\title[]{Spatially resolved resonant tunneling on single atoms in silicon}

\author{B. Voisin$ˆ1$, J. Salfi$ˆ1$, J. Bocquel$ˆ1$, R. Rahman$ˆ2$ and S. Rogge$ˆ1$}

\address{$ˆ1$ Centre for Quantum Computation and Communication Technology, School of Physics, The University of New South Wales, Sydney, NSW 2052, Australia}
\address{$ˆ2$ Purdue University, West Lafayette, IN 47906, USA}
\ead{benoit.voisin@unsw.edu.au}
\vspace{10pt}
%\begin{indented}
%\item[]\today
%\end{indented}

\begin{abstract}
The ability to control single dopants in solid-state devices has opened the way towards reliable quantum computation schemes. In this perspective it is essential to understand the impact of interfaces and electric fields, inherent to address coherent electronic manipulation, on the dopants atomic scale properties. This requires both fine energetic and spatial resolution of the energy spectrum and wave-function, respectively. Here we present an experiment fulfilling both conditions: we perform transport on single donors in silicon close to a vacuum interface using a scanning tunneling microscope (STM) in the single electron tunneling regime. The spatial degrees of freedom of the STM tip provide a versatility allowing a unique understanding of electrostatics. We obtain the absolute energy scale from the thermal broadening of the resonant peaks, allowing to deduce the charging energies of the donors. Finally we use a rate equations model to derive the current in presence of an excited state, highlighting the benefits of the highly tunable vacuum tunnel rates which should be exploited in further experiments. This work provides a general framework to investigate dopant-based systems at the atomic scale.
\end{abstract}

%\submitto{\JPCM}
%\maketitle

\newpage

\section{Introduction}

Solid-state devices have now reached the single-atom level due to the constant improvement of nanofabrication techniques over the last few decades. It is now possible to probe, couple and manipulate single atomic orbitals using a wide range of systems and techniques~\cite{Koenraad2011}. Nevertheless the presence of control electrodes and interfaces, essential to manipulate such quantum orbitals or spin degrees of freedom, can induce strain or quantum confinement which modify the properties of such objects (spectrum, spin states, couplings, lifetimes). Additionally these objects can be very sensitive to their atomic position in the crystal host~\cite{Kane1998, Yakunin2007, Hao2009, Diarra2008}. Silicon represents a prime candidate for future quantum computation schemes due to the extraordinarily long coherence times of donors~\cite{Pla2012,Pla2013, Morton2008}, together with intensive work devoted to their atomically precise positioning in the silicon host~\cite{Fuechsle2012}. However the situation is here more complex because of the peculiar presence of the six degenerate conduction band minima ("valleys")~\cite{Kohn1955, Ramdas1981}. Valleys govern the donor spectrum and properties as they mix in the deep confinement potential of such atoms. The valley mixing, and thus the resulting spectrum, is also perturbed by the local environment~\cite{Koiller2001, Rahman2011b}. For this reason, understanding the complexity around donors, interfaces, electric fields and valleys has become a crucial challenge in the view of donor-based quantum computation~\cite{Zwanenburg2013}.\\~~

Field-effect transistors issued from the microelectronics industry have been natural tools to investigate single donors in silicon~\cite{Hofheinz2006, Sellier2006, Khalafalla2007, Lansbergen2008, PierreM.2010b, Golovach2011, Verduijn2013, Voisin2014}. The well-known Moore's law has accurately predicted the dramatic reduction of their dimensions over the last decades and the sub 10 nm range has now been reached~\cite{Ferain2011}. This allows to electrically address single donors in the channel. At low temperature, in the resonant tunneling regime, it becomes possible to  perform a transport spectroscopy of the donor states~\cite{Lansbergen2008, Roche2012}. However the extreme compactness of such systems from the large scale integration scheme, affording strong couplings to the different electrodes, leads to a very complex 3D electrostatic problem preventing a precise spatial analysis, which is necessary to evaluate the atomic-scale fabrication methodology for donors. On the other hand, a technique like Scanning Tunneling Microscopy (STM) and Spectroscopy (STS) readily offers atomic resolution. STM experiments have been conducted on subsurface single impurities in III-V semiconductors. In particular, the peculiar geometry of cross-sectional STM and the absence of surface states, has allowed to probe and manipulate shallow donors and acceptors~\cite{Feenstra1994, Zheng1994, Kort2001, Mahieu2005, Wijnheijmer2011} as well as deep magnetic ions~\cite{Yakunin2007,Bocquel2013}. In such experiments, the relative position of the dopants and their depth is determined with atomic precision. However the absolute energy scale has usually remained unknown mainly because of tip-induced band bending~\cite{Loth2006, Muennich2013}, even if efforts have been made to convert the bias voltage scale into an absolute energy scale~\cite{Feenstra2003}.
With similar purpose, we show here how we can combine advantages of both transport spectroscopy and STM techniques. We achieve the resonant tunneling regime in a STM/STS setup allowing for an accurate determination of the different energies along with an atomic spatial resolution \cite{Mol2013, Miwa2013, Salfi2014}, a general framework which can be applied to various systems. We perform transport through single donors located in the depletion layer of a highly n-type doped silicon substrate and determine the energy scale from the lineshape of the resonant tunneling onset at 4.2 K. Moreover, varying the tunneling barrier allows to define the electric field at the donor site. The outer tunnel rate can be tuned over more than two orders of magnitude, not possible in transport experiments. A direct application is the determination of the charging energy of such donors close to a vacuum interface. Finally we develop a rate equations model to derive the current in the presence of an excited state, taking into account relaxation \cite{Stoof1996, Bonet2002}. We demonstrate that the ability to tune the different tunnel rates, by varying tip height and sample engineering \cite{Miwa2013}, and taking into account their intrinsic asymmetries, is essential to further investigate the spectrum of single dopant-based systems. 

\section{Sample preparation}

\begin{figure*}
\begin{center}
\includegraphics[width=\textwidth]{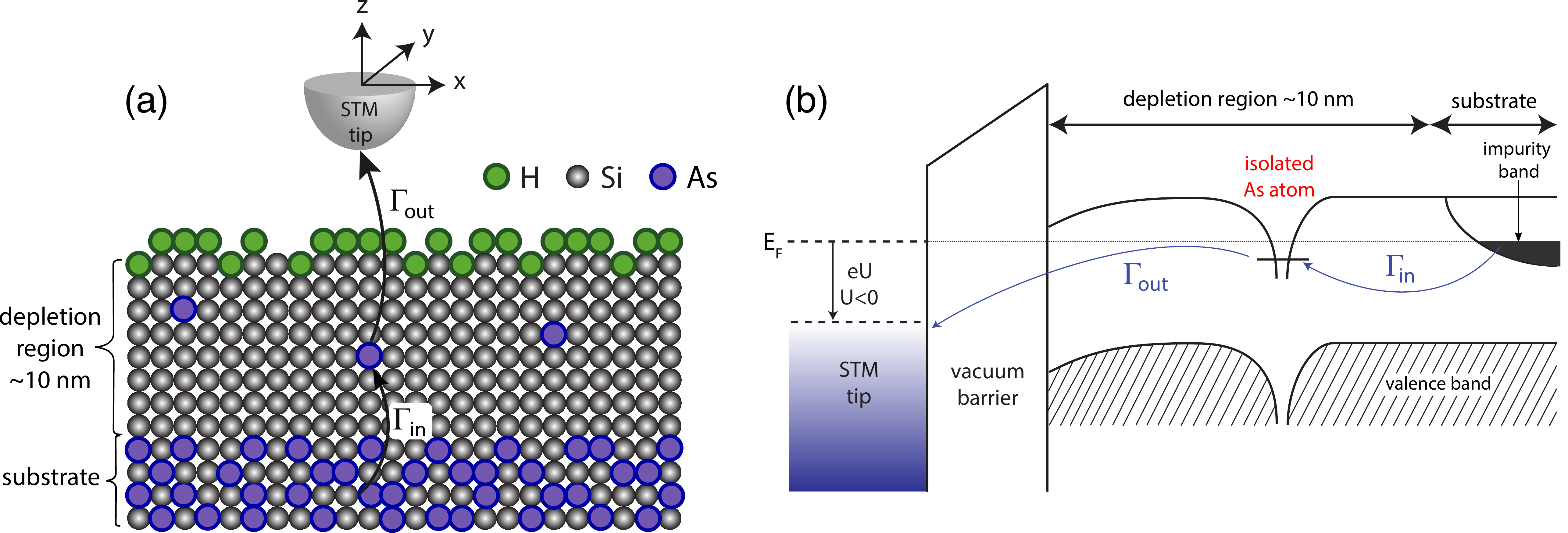}
\caption{(a) Layout of the sample. Resonant tunneling is performed on an isolated As donor located in the depletion layer of a doped silicon wafer. Electrons tunnel from the n-type substrate to the tip. (b) Band diagram in the resonant tunneling conditions. The voltage bias (U) pulls the donor ground state in resonance, between the Fermi levels of the sample's impurity band and of the tip. The donor is located between two tunnel barriers with rates  $\mathrm{\Gamma_{in}}$ and $\mathrm{\Gamma_{out}}$.}
\label{fig: layout and diagram}
\end{center}
\end{figure*}

We start with a commercial n-type doped silicon (100) wafer with a resistivity of 0.004-0.001 $\mathrm{\Omega.cm}$, corresponding to an As concentration of about  $\mathrm{3\cdot10^{19}\,cm^{-3}}$. After cleaving and cleaning, the sample is placed in ultra-high vacuum (pressure below $\mathrm{5\cdot10^{-11}}$\,mbar) and degassed at 600$\mathrm{\degree C}$ for 10 h, to remove the last contaminants. The sample is then flashed three times at 1080$\mathrm{\degree C}$ for 10 s. After the final flash, the sample is slowly cooled down to room temperature ensuring a good crystallisation and a $\mathrm{2\times 1}$ surface reconstruction. The flashing step has two notable effects: first it removes the native oxide and secondly it creates a depletion layer at the surface of the silicon \cite{Pitters2012}. In these conditions, the depletion layer (low As concentration) is about 10 nm-thick as shown in Fig.~\ref{fig: layout and diagram}(a). Finally to passivate the surface states, the reconstructed Si surface is hydrogen terminated employing a thermal cracker at a nominal $\mathrm{H_2}$ pressure of  $\mathrm{5\cdot10^{-7}}$\,mbar for 10 mn at 330$\mathrm{\degree C}$~\cite{Boland1990}. After preparation the sample is transfered into a low temperature STM for electrical measurements. All the experimental data in the article were taken at 4.2 K with a base pressure below $\mathrm{2\cdot10^{-11}}$\,mbar.

\section{Formation and electrostatics of the barriers}

The depletion layer created during the flashing step induces a tunnel barrier, with a rate $\mathrm{\Gamma_{in}}$, between the substrate and any of the remaining isolated As atoms located in this depletion layer. The vacuum barrier between a dopant and the STM tip has a tunneling rate $\mathrm{\Gamma_{out}}$. Eventually, as depicted in Fig.~\ref{fig: layout and diagram}(a), such a dopant is confined between two tunnel barriers, similarly to a dopant centred in the channel of a MOSFET. At low temperature the thermal energy $k_BT$ drops below the charging energy $E_C$ (the energy necessary to bind an extra electron to the donor, usually a few tens of meV \cite{Rahman2011a}), as well as below the single level spacing (the energy difference between the ground and the first excited state, given the by the so-called valley-orbit splitting on the order of a few meV \cite{Rahman2011b, Roche2012}). In these conditions transport occurs in the single electron resonant tunneling regime~\cite{Beenakker1991}. We can give a general expression for the tunnel current as a function of the sample bias $U$ following~\cite{Foxman1993}, assuming $\mathrm{\Gamma = \Gamma_{in}+\Gamma_{out} \ll}\,k_BT$:

\begin{equation}
I(U)=[I_0(z)+\Delta_{\Gamma}(U_{thres}-U)]\int_0^{|U|}{cosh^{-2}\left(\frac{\alpha e (U'-|U_{thres}|)}{2k_BT}\right)dU'}
\label{eq:tunnel current}
\end{equation}

$U_{thres}$ is called the threshold voltage and refers to the onset of the resonance. The parameter $\mathrm{\Delta_\Gamma}$ indicates that the triangular (because of the presence of an electric field) vacuum barrier is lowered when the bias is varied towards more negative values, and so the electric field increased. Therefore the tunnel rate is increased and we assume for simplicity a linear dependence of the current with the bias, i.e. constant $\mathrm{\Delta_\Gamma}$, as the first order correction to the barrier lowering. The parameter $\alpha$ expresses the potential variation at the donor site as a function of the bias: it is the key to convert the sample bias into an absolute energy scale on the donor, independently of the initial tip-induced band bending at zero bias. Finally, $I_0(z)$ represents the amplitude of the current step and varies as a function of the tip height as developed below (note that $I_0$ also depends on the plane coordinates $x$ and $y$). Under these assumptions, a characteristic feature of the resonant tunneling regime is the exponential increase of the current at the onset of the resonant peak when the ground state of the donor is pulled in resonance with, and further goes below, the Fermi level defined by the conduction impurity band of the substrate (see Fig.~\ref{fig: layout and diagram}(b)). ~~\\

\begin{figure*}
\begin{center}
\includegraphics[width=8.5cm]{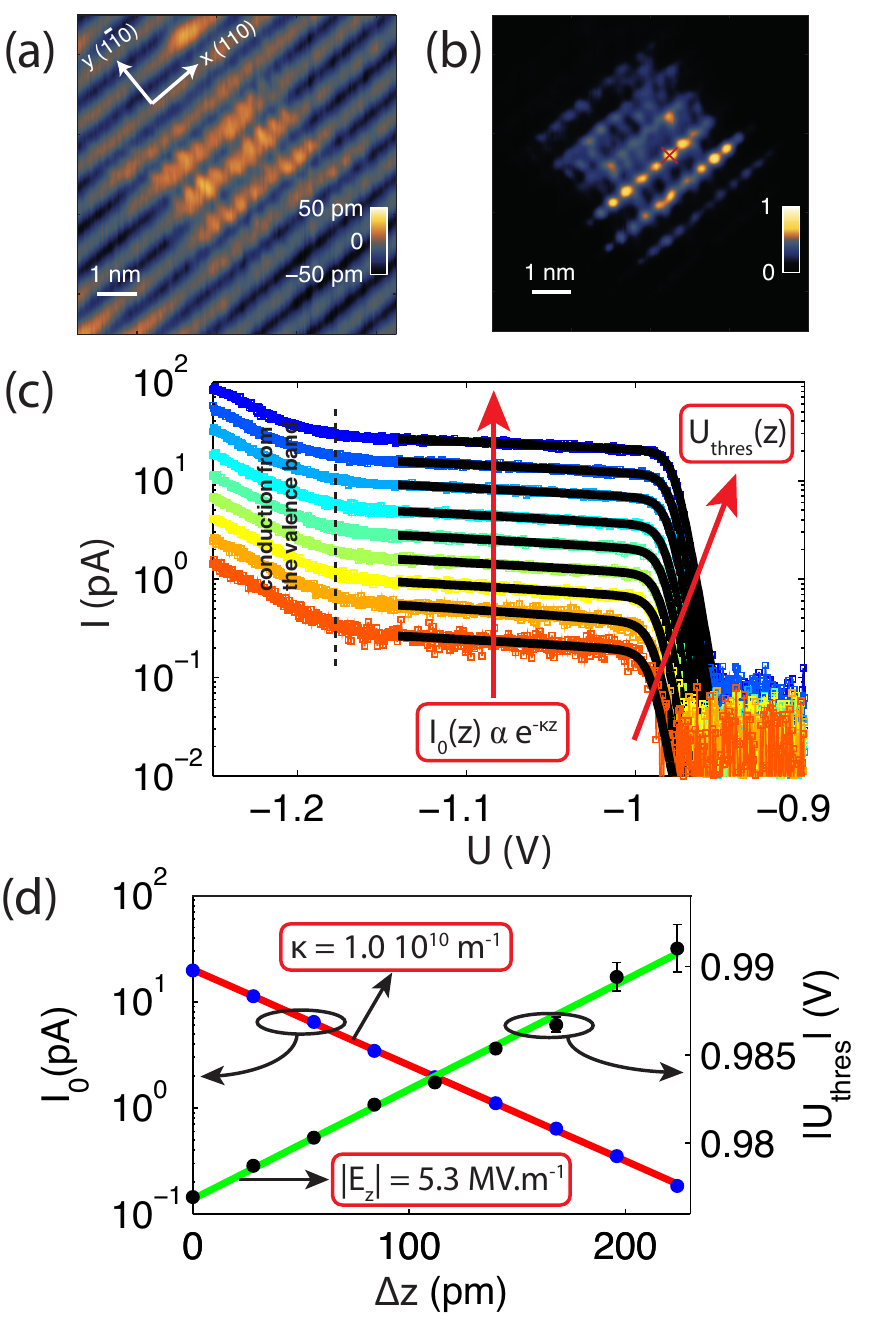}
\caption{Donor 1. (a,b) Two-pass scan over a single donor. First a filled-states topography is performed, shown on (a) (set-point: $U$=-1.25 V, $I$=-100 pA) where the donor is barely visible due to the direct tunneling from the valence band to the tip. Secondly a current measurement (the topography, recorded on (a), is now played with a $z$-offset of 100 pm towards the surface. $U$=-1.0 V) is performed, shown on (b). The voltage is set in the band gap such that only resonant tunneling through the donor is measured, allowing to directly relate the tunnel current to the donor's wave-function. (c) $I$($U$) curves taken above the donor (at the red cross on (b)) at 9 different tip heights (28 pm-step increase). The exponential increase of current as the tip gets closer to the surface proves that  $\mathrm{\Gamma_{in} \gg \Gamma_{out}}$. (d) The fit of the amplitude of the current as a function of $z$ (left axis) allows to extract the vacuum decay length. The variation in the resonance's threshold (right axis) is related to the electric field at the donor site.}
\label{fig: electrostat}
\end{center}
\end{figure*}

We have developed in~\cite{Salfi2014} a procedure to identify single donors using a two-pass scan, summarised in Fig.~\ref{fig: electrostat}(a,b). First, in (a), a filled-states topography is performed ($U$=-1.25 V, $I$=-100 pA). Dopants appear in this image as bright protusions caused by the overlap between the evanescent tail of the donor's wavefunction in the vacuum and the tip's wavefunction.~\cite{Sinthiptharakoon2014}. Its signal is weak compared to the direct tunneling from the valence band to the tip. Immediately after, a second pass records the current while the topography of the first one is played (minus a constant offset to enhance the tunnel current).  The voltage of this second pass is set below the onset of the valence band where only resonant tunneling through donors contributes to the tunnel current (b). Fig.~\ref{fig: electrostat}(c) shows nine $I$($U$) curves taken over this donor for tip heights varying with 28 pm-step increase, fitted using eq.\,(\ref{eq:tunnel current}). The $z$-displacement was carefully calibrated using the well-known distance between atomic planes. Here, several observations are noteworthy. First we observe a exponential increase of the current at the onset of the resonance, demonstrating thereby the validity of the assumption $\mathrm{\Gamma} \ll k_BT$. Secondly we observe an exponential increase of the tunnel current $\mathrm{I_0(z)}$ when the tip gets closer to the silicon surface (see Fig.~\ref{fig: electrostat}(d), left axis). This evolution corresponds to the situation where $\mathrm{ \Gamma_{out} \ll \Gamma_{in}}$, as expected because of the presence of the vacuum. Moreover this situation also corresponds to a classical STM experiment allowing to directly convert the tunnel current measured in Fig.~\ref{fig: electrostat}(b) into the electronic probability distribution at the silicon surface, that is imaging in real space the donor's wave-function. This has been applied in~\cite{Salfi2014} to directly image the valley interferences occurring in the vicinity of such a donor. We can extract the inverse of the vacuum decay length from this exponential behaviour, plotted in Fig.~\ref{fig: electrostat}(d), left axis,  giving $\kappa$=1.0$\mathrm{\cdot 10^{10}\,m^{-1}}$, which reflects the expected vacuum barrier height of about 4.0 eV for shallow donor states. Finally the evolution of $U_{thres}$ as a function of $z$, plotted in Fig. \ref{fig: electrostat}(d), right axis, can be related to the electric field felt by the donor, following the relation derived in~\cite{Salfi2014}:

\begin{equation}
E_z=\frac{1}{\epsilon_{Si}}\frac{d|U_{thres}|}{dz}
\label{eq:elec field}
\end{equation}

We obtain a linear dependence of $U_{thres}$ with $z$ corresponding to an electric field of 5.3 $\mathrm{MV.m^{-1}}$ on the donor. An ionization at low electric field, i.e. close to flat-band, is expected as the energy difference between the conduction band and the Fermi level of a doped substrate equals the ionization energy of a single donor. The flat-band voltage $U_{FB}$ can vary from donor to donor due to differences in the tip and sample work functions but usually found around -0.8 V. Due to variations in the $alpha$ parameter, from the donor's electrostatic environement, as well as in the donor's ionization energy (a few meV~\cite{Salfi2014}), we usually found single donor's ionization occuring between -0.7 V and -1 V. ~~\\

The electrostatics around the donor can be accessed by understanding the dependence of the tunnel current on the vacuum barrier, and to the ability to move the tip in the three spatial directions. The absolute energy scale is deduced from the level arm parameter $\alpha$ obtained in resonant tunneling regime. We obtained an average $\alpha$ of 0.08, with no significant variation as a function of the tip height. We have assumed a simple 1D electrostatic model, since the STM ensures to measure a donors isolated from any defect or cluster. We also obtained for the parameter $\mathrm{\Delta_\Gamma}$ to exponentially vary between $\mathrm{3.4\,MHz.V^{-1}}$ and $\mathrm{2.6\cdot10^{2}\, MHz.V^{-1}}$ as the tip is brought closer to the surface. Finally we can give a rough estimate for $\mathrm{\Gamma_{in}}$, since it should be smaller than the thermal broadening and larger than the maximum tunnel out rate (about 1 nA). This gives $\mathrm{6\,GHz\,<\,\Gamma_{in}\,<\,360\,GHz}$, below the range where inelastic cotunneling or Kondo processes occur and would drastically modify the physics of our system.\\~~

The STM current comes from the overlap between the tip orbital wave-function and the evanescent tail of the donor's wave-function in the vacuum. Thus one could expect an exponential decrease of the tunnel current when the donor gets deeper in the silicon. However this exponential dependence should only occur when the donor's depth exceeds several nanometers. For shallower donors, the quantum confinement of the vacuum interface is more important than that of the dielectric-screened Coulomb potential. In this regime the surface repels the wave-function towards the substrate, and the donor orbital probability density at the surface remains almost constant. Thus we are able to measure donors located up to 2 to 3 Bohr radii (the Bohr radius equals 2.2 nm for As in Si) below the silicon surface. The donor's depth can be independently obtained by fitting the shift of the conduction band's threshold due to the Coulomb potential of the positively charged donor at positive bias, or by comparing the current images to tight-binding simulations, with the wave-functions exhibiting a succession of type A/type B (see ref~\cite{Salfi2014}) and different lateral extents when the donor's depth varies by atomic planes. Here we have ignored for simplicity the various contributions of tip orbitals~\cite{Chen1990} and the local density of states of the substrate~\cite{Mottonen2010, PierreM.2010b}.

\section{Transport characteristics}
Having established a single electron transport regime for single-dopant measurements in scanning tunneling spectroscopy in Fig.~\ref{fig: electrostat}, several aspects of the observed tunneling spectrum can be understood. In the following section we focus on the charging energy required to add another electron to the system, and on the impact of the asymmetric tunnel rates on the tunneling through excited states.

\subsection{Charging states}
We now increase the bias until the point where two electrons can be bound and tunnel through the donor. According to the diagram of Fig.~\ref{fig: layout and diagram}(b), two new transport channels can open when the bias $|U|$ is increased (with $U<0$): first electrons can directly tunnel from the valence band to the tip and secondly a potential well is created at the vacuum interface as the conduction band is pulled below the Fermi level of the substrate. Therefore this well can be filled with an electron resulting in a so-called tip-induced state. If the valence band tunneling presents only a moderate interest, the tip-induced state is a very rich system as it can hybridize with the donor states. A detailed study goes beyond the scope of this article and we will here only give qualitative arguments to highlight this topic.~~\\

\begin{figure}
\begin{center}
\includegraphics[width=8.5cm]{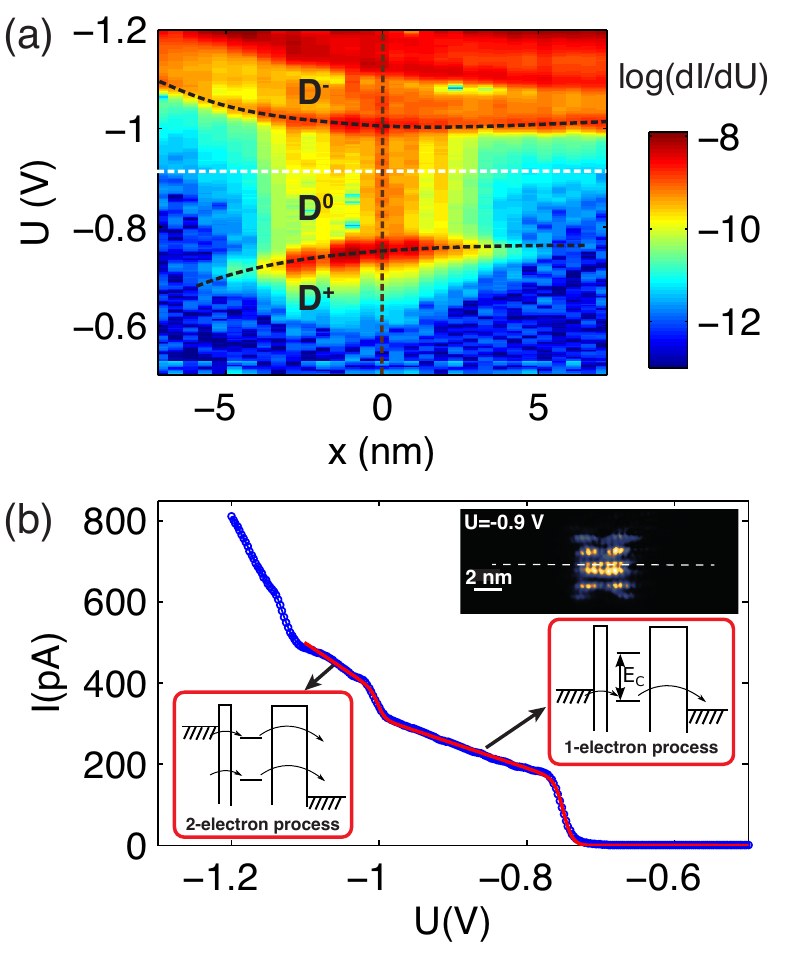}
\caption{Donor 2. (a) Spectroscopy $I$($U$) taken across a donor. We identify two current steps, highlighted by the two black dotted lines corresponding to the $D^+/D^0$ (lower one) and  $D^0/D^-$ (upper one) transitions. The onset of the valence band is found at large negative bias. (b) Cut taken along the vertical brown dotted line in (a). The fit gives an $\alpha$ parameter for each transition and allows to deduce a charging energy of 17 meV. Inset: current image of donor 2 ($U$=-0.9 V). The white dotted line indicates the spatial direction along which the spectroscopy shown in (a) has been performed.}
\label{fig: Dminus}
\end{center}
\end{figure}

In Fig.~\ref{fig: Dminus}(a), a map of the differential conductance (d$I$/d$U$, obtained numerically) vs. the bias is plotted for different points along a line crossing transversly a donor (donor 2, see white dotted line in the inset of Fig.~\ref{fig: Dminus}(b)). The resonant tunneling peak that shifts from -1.0 V to -1.1 V (indicated by the upper black line) is found for all tip positions. Together with the observation that the tip field is attractive at this bias $U<U_FB$, these resonances can be readily ascribed to a tip-induced state. An $I$($U$) curve, taken at $x$=0 along the vertical brown dotted line (i.e. right above the donor), is plotted in Fig.~\ref{fig: Dminus}(b): we observe two current steps corresponding to the transition $D^+/D^0$ (1-electron process, lower black line) and $D^0/D^-$ (2-electron process, upper black line). We have fitted this curve by summing two independent one-electron I(U) given by eq.~\ref{eq:tunnel current}, with thus two different thresholds, amplitudes and $\alpha$ parameters, getting  $\alpha_{D^0}=0.06$ and $\alpha_{D^-}=0.07$, but using a common $\Delta_\Gamma$ for the two steps. We have computed the charging energy as the energy difference between the two peaks, assuming for simplicity that $\alpha$ is a linear function of the energy: we obtain $E_C$=17 meV. This is a lower value as compared to the bulk (53 meV) but in agreement with reference~\cite{Salfi2014} and with the charging energies observed in MOSFET structures~\cite{Lansbergen2008, Verduijn2013, Voisin2014}, where interfaces, leads (tip, substrate) and electric fields~\cite{Rahman2011a} have a crucial role. We can finally note that the tip-induced state, highlighted by the upper black dotted line in Fig.~\ref{fig: Dminus}(a), exists even away from the donor, and goes towards lower energy (or lower bias) above the donor as a result of the hybridization with the donor. The existence of this tip-induced quantum state, its coupling to the donor and finally the absence of nearby extra features such as vacancies or dangling bonds (which would be easily identified in the topography) undoubtedly rule out the possibility of any defect, dopants dimer or cluster, to account for this two-electron state.

\subsection{Master equation - Probing excited states}

The spectrum of a single donor nearby an interface in silicon is nontrivial due to the 6-fold degenerate Si conduction band minimum, called "valleys". The symmetry-breaking perturbation of a planar interface influences the lifetime of spin and valley-orbit excited states, the energy of the latter, and the hyperfine coupling with the nucleus, to name a few effects. Here we use simple rate equations, in the single electron regime (i.e. maximum 1 electron on the donor) to understand how excited states can be probed in STM experiments~\cite{Stoof1996, Bonet2002}:

\numparts
\begin{eqnarray}
\frac{d\rho_g}{dt}=\dot{\rho_g}=(1-\rho_g-\rho_e)\Gamma_{in}^g-\rho_g\Gamma_{out}^g+W\rho_e \\
\frac{d\rho_e}{dt}=\dot{\rho_e}=(1-\rho_g-\rho_e)\Gamma_{in}^e-\rho_e\Gamma_{out}^e-W\rho_e
\end{eqnarray}
\endnumparts

The term $\rho_g$ (resp. $\rho_e$) represents the average occupation number of the ground (resp. excited) state. Note the asymmetric roles played by $\mathrm{\Gamma_{in}}$ and $\mathrm{\Gamma_{out}}$: $\mathrm{\Gamma_{in}}$ adds an electron from the reservoir on the donor with respect to the total occupation number  $\rho_g+\rho_e$, ensuring a maximum of one electron in the system, fulfilling the single electron regime condition. $\mathrm{\Gamma_{out}}$ empties each state at a rate proportional to its occupation number. W represents a phenomenological relaxation rate from the excited state to the ground state.

We can easily derive the stationnary regime solutions $\rho_g^{stat}$,  $\rho_e^{stat}$ by solving $\dot{\rho_g}=\dot{\rho_e}=0$ and then compute the current $I^{g+e}$ flowing through the donor:

\numparts
\begin{eqnarray}
\rho_g^{stat}=\frac{1+\frac{W}{\Gamma_{out}^e}(1+\Gamma_{in}^e/\Gamma_{in}^g)}{(1+\frac{W}{\Gamma_{out}^e})(1+\frac{\Gamma_{out}^g}{\Gamma_{in}^g})+\frac{\Gamma_{in}^e}{\Gamma_{in}^g}(\frac{W+\Gamma_{out}^g}{\Gamma_{out}^e})} \\
\rho_e^{stat}=\frac{1}{1+\frac{W}{\Gamma_{out}^g}+(1+\frac{\Gamma_{in}^g}{\Gamma_{out}^g})(\frac{W+\Gamma_{out}^e}{\Gamma_{in}^e})} \\
I^{g+e}=e\rho_g^{stat}\Gamma_{out}^g+e\rho_e^{stat}\Gamma_{out}^e
\end{eqnarray}
\endnumparts

\begin{figure}
\begin{center}
\includegraphics[width=8.5cm]{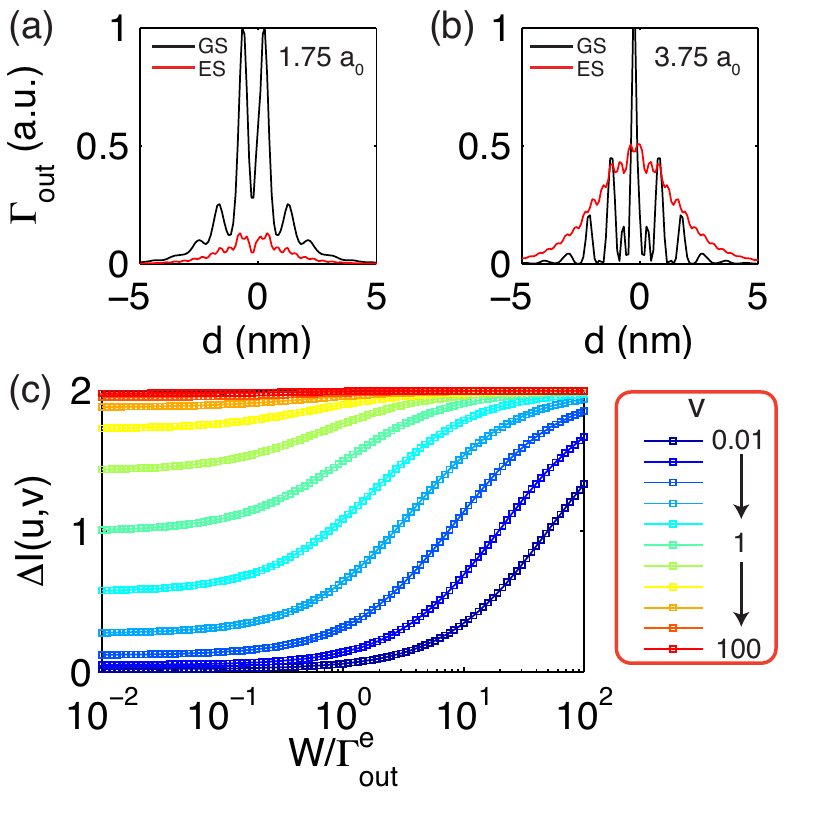}
\caption{(a), (b) $\mathrm{\Gamma_{out}}$ (in arbitrary units) computed for the ground state (GS, in black) and excited state (ES, in red) states of a donor located at $\mathrm{1.75\,a_0}$ (a) and $\mathrm{3.75\,a_0}$ (b) below the surface. This shows the possible differences in the brightnesses ratio  $\mathrm{\Gamma_{out}^e/ \Gamma_{out}^g}$. (c) $\mathrm{\Delta} I$ calculated for various ($u$, $v$) both varying from 0.01 to 100. A decrease of current is observed when the excited state enters the bias window if $u, v \ll 1$. }
\label{fig: simul}
\end{center}
\end{figure}

These expressions are very general and can be applied to any system. Here we give some insights on the different situations which can occur in our STM setup. To simplify these expressions we assume $\mathrm{\Gamma_{in}^g,\Gamma_{in}^e \gg W, \Gamma_{out}^{g,e}}$ , limit which indeed corresponds to our experimental situation.  As previously mentioned, the vacuum tunnel barrier for electrons tunneling to the tip is much slower than the one formed by the depleted region between the reservoir and the donor, and therefore equation (4c) becomes independent of  $\mathrm{\Gamma_{in}}$. We finally derive our quantity of interest which is the variation in the tunnel current when the excited state enter the bias window. For this purpose we define a function $\Delta I$ which reads as follows:

\begin{eqnarray}
\Delta I(u,v)= \frac{I^{g+e}}{I^g}=\frac{2v(1+u)}{1+v(1+2u)}
\end{eqnarray}

with $u=\mathrm{W/\Gamma_{out}^e}$ and $v=\mathrm{\Gamma_{out}^e/\Gamma_{out}^g}$ two dimensionless parameters and $I^g$ the current when only the ground state is in the bias window (thus $I_g=e\mathrm{\Gamma_{out}^g}$). As mentioned above, this formalism is general: the excited state could be for instance a valley or an orbital state, with usually $\mathrm{W \gg \Gamma_{out}}$, or a spin excited state, with possibly $\mathrm{W \ll \Gamma_{out}}$. As shown in Fig.~\ref{fig: electrostat}(c), we are able to tune the outer tunnel rate over three orders of magnitude, from around 60 MHz (corresponding to a tunnel current of 100 fA, noise floor) to around 6 GHz (1 nA, higher currents dehydrogenate the surface), such that the tunnel-out rate could cross the relaxation rate in this range. Secondly we show in Fig.~\ref{fig: simul}(a) and (b) two examples of outer tunnel rates, normalised to the maximum value of the overlap betwen the tip and both of these wave-functions, for the ground and the first valley excited states of a single donor at two different depths: $\mathrm{1.75\,a_0}$ for (a) and $\mathrm{3.75\,a_0}$ for (b), $\mathrm{a_0=0.543}$ nm is the Si lattice constant. These tunnel rates are spatially resolved along a line crossing the donor wave-function at the surface (i.e. along $x$). They have been obtained by computing the density probability of each state at the surface, with the wave-functions being calculated by tight-binding simulations~\cite{Salfi2014, Klimeck2007, Rahman2007}. We observe in (a) that the ground state, plotted in black, is much brighter than the first excited state, plotted in red, which is not the case for the second depth (b). Other depths (not shown) demonstrate even stronger asymmetries without a conclusive trend. Again this shows the sensitivity of the donor properties at the atomic scale due to the presence of valleys, whose coherent populations are modified by the presence of the surface, and are associated with rapid modulation of donor's electronic probability density. Current investigations, concerning single donors excited states, two-electron states or two-donor correlations (exchange for instance), shall give further insights in the role played by silicon valleys on donors' properties.~~\\

To summarize this work on rate equations and the possibility it offers in our STM experiments, we have plotted in Fig.~\ref{fig: simul}(c) the function $\mathrm{\Delta} I$ as a function of $u=\mathrm{W/\Gamma_{out}^e}$ varying from 0.01 to 100, for various $v=\mathrm{\Gamma_{out}^e/\Gamma_{out}^g}$ also varying in the same range. We can notably expect a decrease of current (i.e. $\mathrm{\Delta} I(u,v)<1$) if $u \ll 1$ (slow relaxation) and $v \ll 1$ (slow excited outer rate). This can be understood as a blocked situation where an electron gets trapped in the excited state with both low probabilities to relax and to tunnel to the tip. More conventionally, $\mathrm{\Delta} I(u,v) > 1$ is obtained for $u\ll 1$ and $v \gg 1$ (fast excited out rate). Indeed, low energy single-hole excited states were found in recent STM experiments with acceptors in silicon~\cite{Mol2013}. For donors, the excited state energies can well exceed the thermal energy at 4.2 K. From the preceding discussion we can expect an interplay between tunnel-out rates, relaxation rates, and energies, in determining the influence of excited states on the tunneling current. The spatial dependence of the tunnel rates should be combined with sample engineering~\cite{Miwa2013} to further tackle excited states in future STM experiments.

\section{Conclusion}
In conclusion, we have described a scheme for quantitative single electron transport experiments performed with spatial resolution at low temperatures, using cryogenic scanning tunneling spectroscopy. We have demonstrated the combination of atomic spatial resolution, required to investigate the properties of single donors close to an interface, and of the determination of the absolute enegy scale, essential to obtain addition~\cite{Salfi2014} and excited state energies~\cite{Mol2013}. The method makes use of a level arm parameter extracted from the analysis of the single-electron transport line-shape, which is given by the temperature of a reservoir in the substrate. Some new insights into the electrostatics, based on tip-height dependence measurements of single-dopant transport spectrum, have also been presented. Furthermore we have developed a simple rate equations model showing the strong dependence of the tunnel current as a function of the tunable tunnel rates and relaxation rates. This combination of the STM and of this general formalism can be readily applied to dopants in other systems, and therefore allows to investigate the properties of single and multiple dopants in solid-state devices.

\section{Acknowledgments}

This work is supported by the European Commission Future and Emerging Technologies Proactive Project MULTI (317707) and the ARC Centre of Excellence for Quantum Computation and Communication Technology (CE110001027), and in part by the US Army Research Office (W911NF-08-1-0527). This work is part of the research program of the Foundation for Fundamental Research on Matter (FOM), which is part of the Netherlands Organization for Scientific Research (NWO). S.R. acknowledges a Future Fellowship (FT100100589). The use of nanoHUB.org computational resources operated by the Network for Computational Nanotechnology funded by the US National Science Foundation under grant EEC-0228390 is gratefully acknowledged.

\section*{References}

\bibliography{JPCM2014_STM2}
\bibliographystyle{unsrt}

\end{document}